# A circular microphone array with virtual microphones based on acoustics-informed neural networks

Sipei Zhao[a)] and Fei Ma

Centre for Audio, Acoustics and Vibration, Faculty of Engineering and IT, University of Technology Sydney

**Abstract**

Acoustic beamforming aims to focus acoustic signals to a specific direction and suppress undesirable interferences from other directions. Despite its flexibility and steerability, beamforming with circular microphone arrays suffers from significant performance degradation at frequencies corresponding to zeros of the Bessel functions. To conquer this constraint, baffled or concentric circular microphone arrays have been studied; however, the former needs a bulky baffle that interferes with the original sound field whereas the latter requires more microphones that increase the complexity and cost, both of which are undesirable in practical applications. To tackle this challenge, this paper proposes a circular microphone array equipped with virtual microphones, which resolves the performance degradation commonly associated with circular microphone arrays without resorting to physical modifications. The sound pressures at the virtual microphones are predicted from those measured by the physical microphones based on an acoustics-informed neural network, and then the sound pressures measured by the physical microphones and those predicted at the virtual microphones are integrated to design the beamformer. Experimental results demonstrate that the proposed approach not only eliminates the performance degradation but also suppresses spatial aliasing at high frequencies, thereby underscoring its promising potential.

a) Electronic mail: Sipei.Zhao@uts.edu.au

## I. INTRODUCTION

Beamforming with Circular Microphone Arrays (CMAs) aims to localize sound sources and extracts sound signals from a certain spatial direction, which has been used in various applications such as speech communications (Huang et al., 2017a), hearing aid devices (Meyer, 2001), environmental noise source localization (Tiana-Roig et al., 2010), and direction of arrival estimation (Pavlidi et al., 2013) etc.

The CMA is advantageous over linear arrays because it can localize sound signals over 360° and avoid azimuth ambiguity. Therefore, the past two decades have witnessed tremendous research attention on the CMA and various design methods have been studied. Meyer (2001) investigated the performance of a CMA mounted around a rigid sphere based on the phase mode beamforming with either monopole or dipole sensors. Teutsch and Kellermann (2006) proposed to detect and localize acoustic sources based on wavefield decomposition using CMAs. Tiana-Roig et al. (2010) proposed a Circular Harmonic Beamforming (CHB) method, which shows a better angular resolution and a lower side lobe level than the conventional delay-and-sum beamformer although it is more sensitive to sensor noise. Torres et al. (2012) proposed to combine the CHB with time-frequency processing for robust acoustic source localization using CMAs. Different from the above eigen beamforming technique, Benesty et al. (2015) proposed differential beamforming with CMAs by obtaining the spatial filters through solving some linear equations determined by the null constraints from the desired beampattern. Huang et al. (2020) extended this null-constrained method to include symmetric constraints for CMAs to design a continuously steerable differential beamformer. In addition, other methods for beamforming with CMAs, such as the Jacobi-Anger expansion method (Huang et al., 2017a) and the Minimum Mean Square Error (MMSE) method (Wang et al., 2022), have also been explored.

Despite having been broadly studied, one of the major concerns with an open CMA is the inherent deep nulls in the directivity factor and the white noise gain caused by the zeros of the Bessel functions at some frequencies (Huang et al., 2017b). To overcome this inherent problem in CMAs, researchers have proposed to mount the CMA around a baffle, such as a rigid sphere (Meyer, 2001; Tiana-Roig et al., 2011) or an infinite cylinder, either acoustically soft (Teutsch and Kellermann, 2006) or rigid (Tiana-Roig et al., 2010). However, adding a baffle, either a cylinder or sphere, is usually cumbersome and introduces interferences to the original sound field, which may not be desired in practice. Therefore, Concentric Circular Microphone Arrays (CCMAs) have been studied to overcome the deep nulls problem in the CMA, without the need for a baffle. Huang et al. (2017b) proposed a robust concentric circular differential microphone array based on the Jacobi-Anger expansion to overcome the deep nulls problem in the CMA. Wang et al. (2023) proposed a mode matching-based beamformer using concentric circular differential microphone arrays. While these beamforming techniques based on CCMAs can mitigate the deep nulls problems of CMAs, more microphones are needed to create a CCMA, making the system more complicated and expensive.

In an alternative approach, this paper proposes a circular microphone array with virtual microphones (CMA-VM) based on recent advances in Physics-Informed Neural Networks (PINNs). In the past few years, PINNs have been a hot research topic in various fields (Cai et al., 2021a, 2021b; Misyris et al., 2020); as opposed to the pure data-driven machine/deep learning techniques, the PINN incorporates physical laws into the design and optimization of a neural network to improve its generalization capabilities (Karniadakis et al., 2021). Recently, PINNs have also been applied in acoustics for various applications. Olivieri et al. (2021) applied a PINN for nearfield acoustic holography; Kahana et al. (2023) used a PINN to locate acoustic sources in a waveguide with high accuracy; Ma et al. (2023) applied a PINN for spatial upsampling of head-related transfer functions; Pezzoli et al. (2023) and Karakonstantis et al. (2024) used a PINN to reconstruct room impulse responses in the time domain and found that the PINN outperforms other methods in reconstructing the early part of the room impulse responses; Ma et al. (2024) proposed a compact Acoustics-Informed Neural network (AINN) to reconstruct a spatial sound field in different room conditions. Despite the above studies, however, PINN has not been used in acoustic beamforming, to the best of the authors' knowledge.

In this paper, based on the recently proposed AINN by the authors (Ma et al., 2024), the CMA-VM approach, which overcomes the deep nulls problem in the CMA yet does not need a baffle or additional microphones to create a CCMA, is proposed. A circular array of physical microphones is used to measure the sound pressures along a ring, based on which the sound pressures at the virtual microphones along a concentric ring are estimated using a compact AINN. The measured sound pressures by the physical microphones and the estimated sound pressures at the virtual microphones are then combined to design the beamformer. By using the AINN-based virtual microphones, the deep nulls problem in the conventional CMA are tackled without introducing extra physical microphones or a cylindrical/spherical baffle. In addition, by using more virtual microphones than the original physical microphones, spatial aliasing effects at higher frequencies can also be suppressed. Experiments are carried out in a Hemi-Anechoic chamber to validate the proposed approach.

## II. CONCENTRIC CIRCULAR MICROPHONE ARRAY

### A. Problem formulation

As illustrated in FIG. 1, a typical CCMA consists of $Q$ rings, each with $M_q$ ($q$ = 1, 2, …, $M$) microphones. Therefore, the total number of microphones used in the CCMA is $M = \sum_{q=1}^{Q} M_q$. For notational simplicity and without loss of generality, the centre of the CCMA is assumed to coincide with the origin of the coordinate system and the azimuth angles are measured anti-clockwise from the positive $x$-axis, as depicted in FIG. 1. The radius of the $q$-th ring is $r_q$ and the angular position of the $m$-th microphone on the $q$-th ring is $\varphi_{q,m}$. The microphones on each ring are assumed to be uniformly distributed, hence (Wang et al., 2024)

$$\varphi_{p,m} = \varphi_{p,1} + (m-1)\frac{2\pi}{M_p}, \tag{1}$$

where $\varphi_{q,1}$ denotes the angular position of the first microphone on the $q$-th ring.

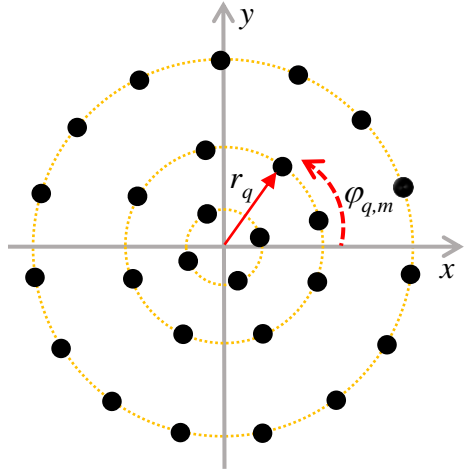

FIG. 1. Illustration of the Concentric Circular Microphone Array (CCMA).

Assuming that a plane wave at frequency $f$ with an amplitude of $A(f)$ impinges on the CCMA from the direction of $\theta$, the sound pressure at the $m$-th microphone in the $q$-th ring can be expressed as

$$p(kr_q, \varphi_{q,m}) = A(f)e^{jkr_q \cos(\theta - \varphi_{q,m})}, \tag{2}$$

where $j$ is the imaginary unit and $k = 2\pi f/c$ denotes the wave number and $c$ = 340 m/s the speed of sound. To filter out the sound signals from a particular direction, a complex weight, $h^*_{q,m}(f)$, is applied to the sound pressure measured by each microphone, and the weighted outputs are superposed to form the array response, i.e.,

$$y(f,\theta) = \sum_{q=1}^{Q}\sum_{m=1}^{M_q} h_{q,m}^*(f) p(kr_q, \varphi_{r,q}), \tag{3}$$

where the superscript $^*$ denotes the complex conjugate operator. Substituting Eq. (2) into Eq. (3) yields

$$y(f,\theta) = A(f)\sum_{q=1}^{Q}\sum_{m=1}^{M_q} h_{q,m}^*(f) e^{jkr_q \cos(\theta - \varphi_{q,m})}. \tag{4}$$

Equation (4) shows that the output of the array is a function of the angle of the incident plane wave. The objective of a beamforming algorithm is to optimize the weights **h**(*f*) so that the microphone array picks up sound signals from a certain look direction, $\theta_s$, while suppressing the sound signals from all other directions. Therefore, the ideal beamformer response is

$$b_{\text{ideal}}(f,\theta) = A(f)\delta(\theta - \theta_s), \tag{5}$$

where $\delta(\cdot)$ is the Kronecker delta function. This is can be expanded in Fourier series as (Tiana-Roig et al., 2010)

$$b_{\text{ideal}}(f,\theta) = A(f)\sum_{n=-\infty}^{\infty} e^{jn(\theta - \theta_s)}. \tag{6}$$

**B. Beamformer design**

To approximate the beamformer response in Eq. (4) to the ideal response in Eq. (6), the exponential term in Eq. (2) is expanded into circular harmonics according to the Jacobi-Anger expansion, i.e., (Huang et al., 2017a)

$$e^{jkr_q \cos(\theta - \varphi_{q,m})} = \sum_{n=-\infty}^{\infty} \gamma_{n,q} e^{jn(\theta - \varphi_{q,m})}, \tag{7}$$

where $\gamma_{n,q} = j^n J_n(kr_q)$ and $J_n(\cdot)$ denotes the *n*-th order Bessel function of the first kind. Substituting Eq. (7) into Eq. (3), one obtains

$$\begin{aligned} y(f,\theta) &= A(f)\sum_{q=1}^{Q}\sum_{m=1}^{M_q} h_{q,m}^*(f) \sum_{n=-\infty}^{\infty} \gamma_{n,q} e^{jn(\theta - \varphi_{q,m})} \\ &= A(f)\sum_{n=-\infty}^{\infty} e^{jn\theta} \sum_{q=1}^{Q} \gamma_{n,q} \sum_{m=1}^{M_q} h_{q,m}^*(f) e^{-jn\varphi_{q,m}} \end{aligned}. \tag{8}$$

In principle, the array response in Eq. (8) can approach the ideal beamformer response in Eq. (6) by equating the corresponding coefficients of the infinite number of circular harmonics. In practical implementations, however, the infinite series must be truncated to a certain order due to a limited number of microphones used in each ring of the concentric array. As a rule of thumb, $N_q \approx kr_q$ is usually used as the first approximation due to the fact that the amplitude of the Bessel functions is small when the order of the Bessel function, *n*, exceeds its argument, $kr_q$ (Tiana-Roig et al., 2010). On the other hand, to avoid spatial aliasing, the highest order should be smaller than half the number of microphones on each ring, i.e., $N_q < M_q/2$. Therefore, $N_q = \min(\lceil kr_q \rceil, M_q/2 - 1)$ is chosen for each ring. The maximum order of the concentric array is determined by the maximum order of each ring, i.e., $N = \max(N_1, ..., N_q, ..., N_Q)$. Therefore, the array output in Eq. (8) is approximated as

$$y(f,\theta) \approx A(f) \sum_{n=-N}^{N} e^{jn\theta} \sum_{q=1}^{Q} \gamma_{n,q} \sum_{m=1}^{M_q} h_{q,m}^*(f) e^{-jn\varphi_{q,m}}. \tag{9}$$

Similarly, the ideal beamformer response in Eq. (6) is also truncated to a finite number of circular harmonics, i.e.,

$$b_N(f,\theta) = A(f) \sum_{n=-N}^{N} e^{jn(\theta-\theta_s)}. \tag{10}$$

By equating Eqs. (9) and (10), one obtains

$$j^n \sum_{q=1}^{Q} J_n(kr_q) \sum_{m=1}^{M_q} h_{q,m}^*(f) e^{-jn\varphi_{q,m}} = e^{-jn\theta_s} \tag{11}$$

for $n \in [-N, N]$, which can be written in vector form as

$$j^n \sum_{q=1}^{Q} \mathbf{h}_q^H(f) \boldsymbol{\psi}_{n,q} J_n(kr_q) = e^{-jn\theta_s}, \tag{12}$$

where

$$\mathbf{h}_q(f) = \begin{bmatrix} h_{q,1}(f) & \cdots & h_{q,m}(f) & \cdots & h_{q,M_q}(f) \end{bmatrix}^T \tag{13}$$

and

$$\boldsymbol{\psi}_{n,q} = \begin{bmatrix} e^{-jn\varphi_{q,1}} & \cdots & e^{-jn\varphi_{q,m}} & \cdots & e^{-jn\varphi_{q,M_q}} \end{bmatrix}^T \tag{14}$$

are vectors of length $M_q$. Equation (12) can be further simplified as

$$j^n \mathbf{h}^H(f) \boldsymbol{\psi}_n(f) = e^{-jn\theta_s}, \tag{15}$$

where

$$\mathbf{h}(f) = \begin{bmatrix} \mathbf{h}_1^T(f) & \cdots & \mathbf{h}_q^T(f) & \cdots & \mathbf{h}_Q^T(f) \end{bmatrix}^T \tag{16}$$

and

$$\boldsymbol{\psi}_n(f) = \begin{bmatrix} J_n(kr_1)\boldsymbol{\psi}_{n,1}^T & \cdots & J_n(kr_q)\boldsymbol{\psi}_{n,q}^T & \cdots & J_n(kr_Q)\boldsymbol{\psi}_{n,Q}^T \end{bmatrix}^T \tag{17}$$

are vectors of length $M$.

Taking conjugate for both sides of Eq. (15) yields

$$\psi_n^H(f)\mathbf{h}(f) = j^n e^{jn\theta_s}. \qquad (18)$$

By stacking Eq. (18) for $n \in [-N, N]$, one obtains

$$\Psi(f)\mathbf{h}(f) = \boldsymbol{\beta}, \qquad (19)$$

where

$$\Psi(f) = \begin{bmatrix} \psi_{-N}^H(f) & \cdots & \psi_n^H(f) & \cdots & \psi_N^H(f) \end{bmatrix}^T \qquad (20)$$

is a $(2N+1) \times M$ matrix that has full-column rank and

$$\boldsymbol{\beta} = \begin{bmatrix} j^{-N} e^{-jN\theta_s} & \cdots & j^n e^{jn\theta_s} & \cdots & j^N e^{jN\theta_s} \end{bmatrix}^T \qquad (21)$$

is a vector of length $(2N+1)$. To avoid spatial aliasing, $M$ is usually no less than $(2N+1)$. When $M > (2N+1)$, Eq. (19) is a underdetermined system and the solution is (Huang et al., 2018)

$$\mathbf{h}(f) = \Psi^H(f)\left[\Psi(f)\Psi^H(f)\right]^{-1}\boldsymbol{\beta}, \qquad (22)$$

When $M = (2N+1)$, $\Psi(f)$ is a square matrix and Eq. (19) is a fully determined system, so the solution simplifies to

$$\mathbf{h}(f) = \Psi^{-1}(f)\boldsymbol{\beta}. \qquad (23)$$

Once the weight vector $\mathbf{h}(f)$ is obtained from Eq. (22) or (23), substituting it to Eq. (3) will generate the beamformer response that is steered to the desired direction $\theta_s$.

The above formulation for the CCMA degenerates to the CMA for $Q = 1$, when all the microphones are placed along a single ring. The CMA needs less microphones but suffers from deep degradation in performance due to the zeros of the Bessel functions at some frequencies; this has been discussed in many references, e.g., (Huang et al., 2017b; Tiana-Roig et al., 2010), and thus is not repeated here for brevity. Although the CCMA can overcome this problem, more microphones are required to form a CCMA, which leads to increased cost and complexity. In the next section, it is proposed to replace some microphones in the CCMA with virtual ones to reduce the number of physical microphones.

## III. VIRTUAL MICROPHONES BASED ON AINN

As depicted in FIG. 2, instead of the conventional CCMA with all physical microphones on each ring, this paper proposes to combine physical microphones on a single ring (black dots) and virtual microphones (green dots) to form a Circular Microphone Array with Virtual Microphones (CMA-VM). The sound pressures at the virtual microphones are estimated from those measured by the physical microphones based on a recently proposed Acoustics-Informed Neural Network (AINN) (Ma et al., 2024). The physical and virtual microphones are then combined to form a CCMA so as to eliminate the deep-nulls problem in the conventional CMA. Therefore, the proposed CMA-VM incorporates the benefits of both the CMA (using fewer physical microphones) and CCMA (showing no noticeable performance degradation) without the need for a baffle.

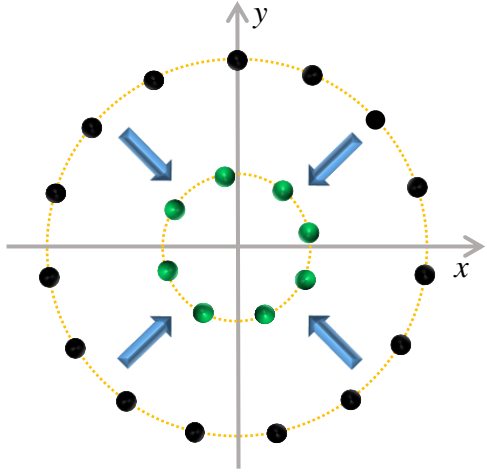

FIG. 2. (Color Online) Illustration of the Circular Microphone Array with Virtual Microphones (CMA-VM).

To estimate the sound pressures at the virtual microphones based on the measured data of the physical microphones, the AINN in FIG. 3 is utilized, where the blue solid circles denote the network neurons. The input to the AINN are the coordinates of the training data, and the output are the corresponding predicted sound pressures. In the traditional purely data-driven technique, only the coordinates of the measurement points (physical microphones) and the corresponding measured data (sound pressures) are used to train the neural network, as depicted by the yellow blocks in FIG. 3. In this case, the loss function to update the network weights is defined as the Mean Square Error (MSE) of the training data, i.e.,

$$\varepsilon_D = \frac{1}{M} \| \mathbf{p}_D - \hat{\mathbf{p}}_D \|^2, \tag{24}$$

where $\|\cdot\|$ denotes the second order norm operator, and $\mathbf{p}_D$ and $\hat{\mathbf{p}}_D$ are column vectors of length $M$ for the sound pressures that are measured by the $M$ physical microphones and that are estimated by the neural network, respectively. The hat on $\hat{\mathbf{p}}_D$ indicates this is an intermediate value duering the network training process.

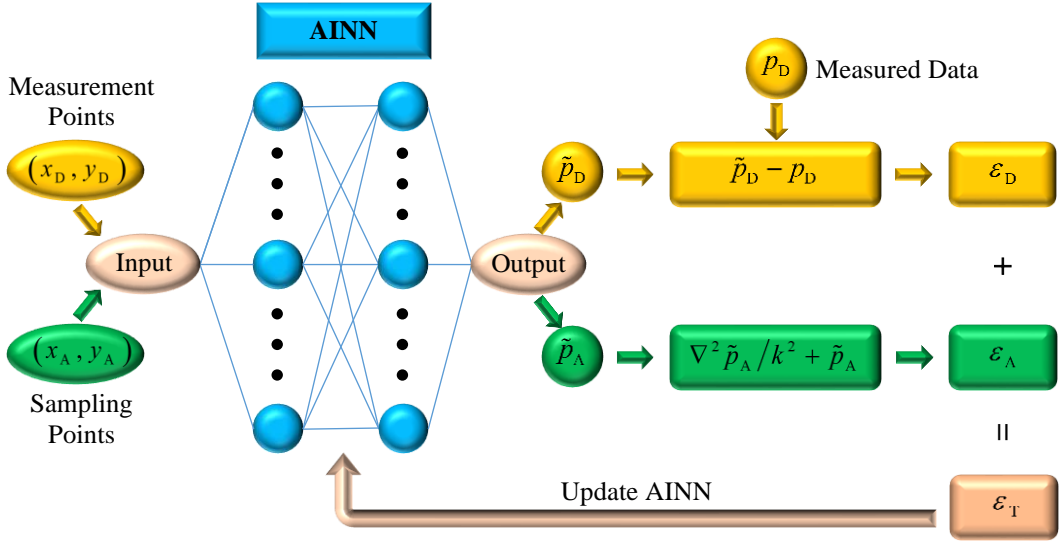

FIG. 3. Diagram of the Acoustics Informed Neural Network (AINN).

In the AINN, in addition to the above data loss function, acoustic laws are also taken into account when optimizing the weights of the neural network. It is well known that the spatial propagation of acoustic waves in linear medium is governed the Helmholtz equation, i.e., (Nelson and Elliott, 1992)

$$\nabla^2 p + k^2 p = 0, \qquad (25)$$

where $\nabla^2$ is the Laplacian operator. To account for this acoustic law, the sound pressure estimated by the neural network is sampled over a spatial region of interest. The coordinates of the sampling points are denoted as $(x_{A,i}, y_{A,i})$ and the corresponding estimated sound pressures are $\hat{p}_{A,i}$ for $i = 1, 2, \ldots, I$, where $I$ is the total number of spatial sampling points. The subscript $_A$ is used to differentiate acoustically sampled points from the measurement points. Therefore, an additional loss function is defined to regulate the neural network, i.e., (Ma et al., 2024)

$$\varepsilon_A = \frac{1}{I} \sum_{i=1}^{I} \left\| \frac{1}{k^2} \nabla^2 \hat{p}_{A,i} + \hat{p}_{A,i} \right\|^2, \qquad (26)$$

where the left-hand side of Eq. (25) has been reformulated in Eq. (26) so that the acoustic loss function has the same physical dimension with the data loss function in Eq. (24). The definition of the acoustic sampling points and the associated acoustic loss function is illustrated by the green blocks in FIG. 3. To consider both the data and the acoustic losses during the training stage, the overall loss function is defined as

$$\varepsilon_T = \varepsilon_D + \varepsilon_A. \qquad (27)$$

The gradients of the overall loss function against the neural network parameters are used to update the network, as depicted by the orange block and arrow in FIG. 3.

In the AINN used in this paper, two hidden layers and one output layer are used and the frequency-dependent number of neurons in each layer is set to $\lceil kr \rceil$, where $\lceil \cdot \rceil$ denotes the ceiling operator and $r$ is the radius of the CMA with physical microphones (Ma et al., 2024). The AINN is implemented with the TensorFlow library and the ADAM algorithm with a learning rate of $10^{-3}$ is used as the optimizer. The training process is set to stop when the data loss does not change over 5000 consecutive epochs. It is noted that the AINN in FIG. 3 is a real-valued network, so the real and imaginary parts of the sound pressures are trained and estimated using two separate AINNs. Our previous results showed that this decoupled-AINN performs better than the coupled-AINN where the real and imaginary parts of the sound pressures are trained and estimated using a single neural network (Ma et al., 2024).

After the AINN is well trained, the sound pressures at the virtual microphones are predicted at the given coordinates, as shown in FIG. 2. Then the sound pressures measured by the physical microphones and that predicted at the virtual microphones are combined and fed to the beamformer. As an example for simplicity, it is assumed that the physical microphones are placed on a ring and the virtual microphones are on another concentric ring to form a dual-ring CMA-VM. In this case, the response of the beamformer can be written as

$$y(f,\theta) = \sum_{m=1}^{M_1} h_{1,m}^*(f) p(kr_1, \varphi_{1,m}) + \sum_{\tilde{m}=1}^{\tilde{M}_2} h_{2,\tilde{m}}^*(f) \tilde{p}(kr_2, \varphi_{2,\tilde{m}}), \tag{28}$$

where $p(kr_1, \varphi_{1,m})$ and $\tilde{p}(kr_2, \varphi_{2,\tilde{m}})$ denote the sound pressures at the $m$-th physical microphone at $(r_1, \varphi_{1,m})$ and the $\tilde{m}$-th virtual microphone at $(r_2, \varphi_{2,\tilde{m}})$, and $h_{1,m}^*(f)$ and $h_{2,\tilde{m}}^*(f)$ denote their corresponding weighting coefficients, respectively. The number of physical microphones $M_1$ does not necessarily equal to the number of the virtual microphone $\tilde{M}_2$; as will be shown in Section IV, using more virtual than physical microphones is able to mitigate the spatial aliasing effect at the high frequency range. The tilde on $\tilde{p}$, $\tilde{M}_2$ and $\tilde{m}$ denotes that these variables are associated with the virtual microphones.

Once the weighting coefficients are designed according to the beamforming algorithm in Section II and the sound pressures at the virtual microphones estimated by the AINN in this section, the beamformer response is calculated according to Eq. (28) and the performance of the CMA, CCMA and CMA-VM is compared in terms of Beampattern, Directivity Index (DI) and White Noise Gain (WNG), respectively. The Beampattern evaluates the response of the beamformer to a plane wave from the direction of $\theta$, hence Eqs. (3) and (28) represent the beampatterns of the CCMA and the CMA-VM, respectively, as a function of frequency $f$ and incident angle $\theta$. The CMA is a special case of the CCMA when $Q = 1$ or equivalently, a special case of the CMA-VM when the virtual microphones are not used (i.e., the second term on the right-hand side of Eq. (28) vanishes). The DI describes the performance of the beamformer in suppressing noise from other directions than the look direction $\theta_s$, and is defined as (Huang et al., 2018)

$$DI(f) = 10\log_{10} \frac{\|y(f, \theta_s)\|^2}{\mathbf{h}^H(f)\mathbf{\Gamma}(f)\mathbf{h}(f)}, \quad (29)$$

where the elements of the $M \times M$ matrix $\mathbf{\Gamma}(f)$ are

$$[\mathbf{\Gamma}(f)]_{ij} = \text{sinc}(k\delta_{ij}), \quad (30)$$

where $\delta_{ij}$ denotes the distance between the $i$-th and $j$-th microphones. The WNG characterizes the robustness of a beamformer and is defined as

$$WNG(f) = 10\log_{10} \frac{\|y(f, \theta_s)\|^2}{\mathbf{h}^H(f)\mathbf{h}(f)}. \quad (31)$$

For a fair comparison between different methods without loss of generality, it is usually assumed the incident wave has a unit amplitude. In the Section IV, experimental results are presented to compare the performance of the proposed CMA-VM method with that of the conventional CMA and CCMA.

## IV. EXPERIMENTS AND DISCUSSIONS

### A. Experimental setup

Experiments were carried out in the UTS Hemi-Anechoic chamber with 50 mm Martini Absorb XHD50 (CSR Martini, Ingleburn, Australia) sound absorbing materials on the ground to reduce acoustics reflections. As shown in FIG. 4, a concentric circular microphone array is placed at the centre of a circular array of 60 loudspeakers. The Genelec 8010A loudspeakers (Iisalmi, Finland) were mounted on a circular truss with a radius of 1.5 m but the radius of the acoustic centres of the loudspeakers is approximately 1.25 m due to the size of the mounting facilities. The DPA 4060 Series Miniature Omnidirectional Microphones (DPA Microphones, Lillerød, Denmark) were uniformly placed along two rings, each with 30 microphones, with the radii of 0.12 m and 0.10 m, respectively. To measure the impulse responses from the loudspeakers to the microphones, a 3 s long log sine sweep signal between 20 Hz and 22 kHz is played back at a sampling rate of 48 kHz for 3 repetitions. More details about the experimental setup and measurement procedures are available in (Zhao et al., 2022). The acoustic transfer functions were obtained from the impulse responses via fast Fourier transform.

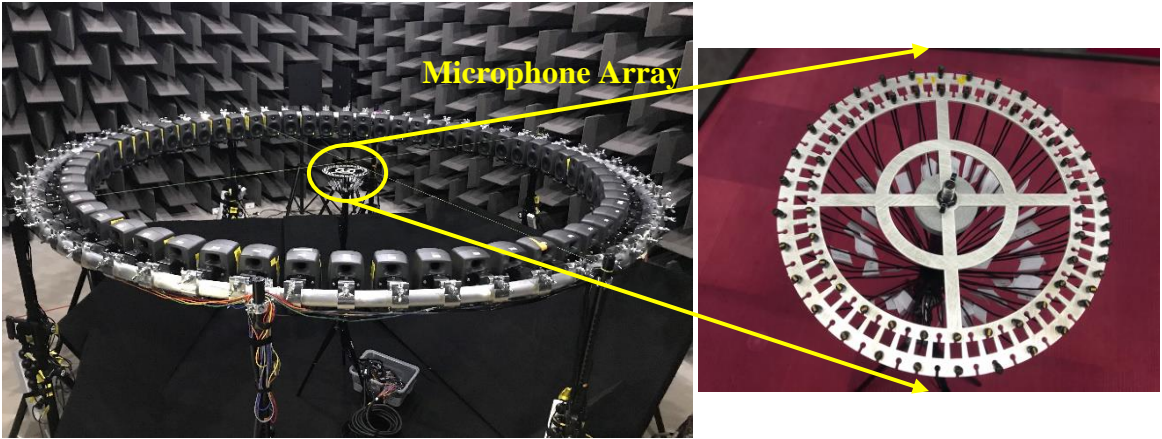

FIG. 4. Photograph of the experimental setup with the loudspeaker array (left) and the microphone array (right). (Not to scale)

**B. Results for 30 physical microphones**

With the measured acoustic transfer functions, the performance of the proposed CMA-VM method is compared with that of the CMA and CCMA. In the first case, $M_1 = 30$ microphones along the outer ring, with a radius of $R_1 = 0.12$ m, are used in the CMA and the corresponding beampattern is shown in FIG. 5(a). It is seen that the beampattern shows a main lobe that narrows with increasing frequency, which is consistent with the results in (Tiana-Roig et al., 2010). In the meantime, the beamformer shows a significant degradation in performance at several frequencies, i.e., 1084 Hz, 1728 Hz, 2316 Hz and 2490 Hz etc. These frequencies correspond to the zeros of the Bessel functions that results in the deep nulls in the DI and WNG, as shown by the black dotted lines in FIG. 6. To remove these deep nulls, an additional $M_2 = 30$ microphones are placed along the inner ring with a radius of $R_2 = 0.10$ m to form a CCMA. The beampattern in FIG. 5(b) for the CCMA shows that the irregularity in the beampattern at those frequencies are eliminated, therefore, the deep nulls in the DI and WNG are removed, as depicted by the blue solid lines in FIG. 6. However, this improvement in performance comes at the expense of double number of physical microphones used in the CCMA, which is undesired in practice due to the induced extra complexity and cost.

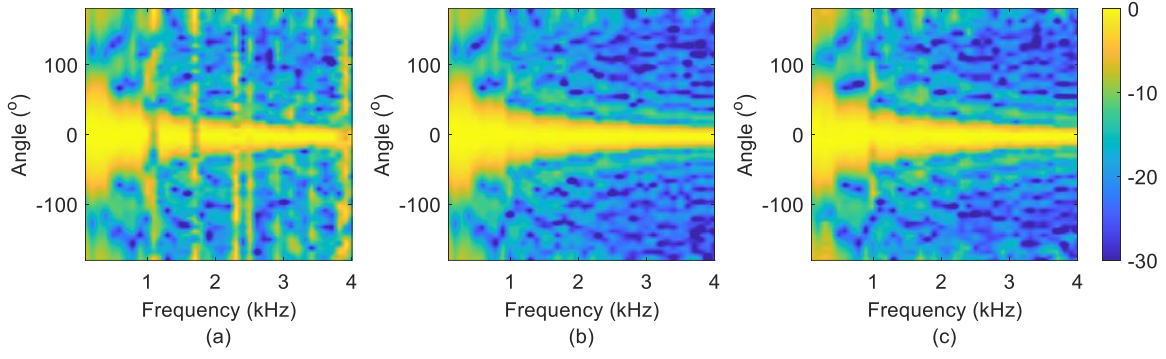

FIG. 5. (Color Online) Beampatterns for (a) the CMA with $R_1 = 0.12$ m and $M_1 = 30$, (b) the CCMA with $R_1 = 0.12$ m, $R_2 = 0.10$ m and $M_1 = M_2 = 30$, (c) the CMA-VM with $R_1 = 0.12$ m, $\tilde{R}_2 = 0.10$ m and $M_1 = \tilde{M}_2 = 30$. The tilde on $\tilde{R}_2$ and $\tilde{M}_2$ denotes that these variables are associated with the virtual instead of physical microphones.

To overcome this limitation, the physical microphones on the inner ring are replaced by virtual microphones to form the proposed CMA-VM. The sound pressures of the $\tilde{M}_2 = 30$ virtual microphones on the ring with a radius of $\tilde{R}_2 = 0.10$ m are predicted from the sound pressures measured by the $M_1 = 30$ microphones physical microphones on the outer ring with a radius of $R_1 = 0.12$ m based on the proposed AINN in Section III. The tilde on $\tilde{M}_2$ and $\tilde{R}_2$ denotes that these variables refer to the virtual microphones instead of physical microphones. The beampattern for the proposed CMA-VM is shown in FIG. 5(c), which presents comparable performance to the CCMA but with only a half number of physical microphones. The deep nulls in the DI and WNG are also eliminated, as evidenced by the red dashed lines in FIG. 6.

The beampatterns for the CMA, CCMA and the proposed CMA-VM methods at the first 4 null frequencies (i.e., $f = 1084$ Hz, 1728 Hz, 2316 Hz and 2490 Hz for a radius of 0.12 m) are compared in FIG. 7. It can be observed that, at these null frequencies, the traditional CMA fails to perform spatial filtering; the CCMA conquers this problem and shows a main lobe that is generally over 10 dB higher than the maximum side lobe level; and the proposed CMA-VM also overcome the impediment and achieves a similar performance to the CCMA except at $f = 1728$ Hz in FIG. 7(b), where the maximum side lobe is approximately -6.0 dB (in comparison to -8.7 dB for the CCMA). This might be due to some unknow measurement noise at this frequency, to be further investigated in future experiments. The above results demonstrate that, using a circular array of 30 physical microphones, the proposed CMA-VM method eliminates the deep nulls of the conventional CMA with the same number and arrangement of microphones and achieves a similar performance to the conventional CMA with 60 physical microphones uniformly placed along two concentric circular rings.

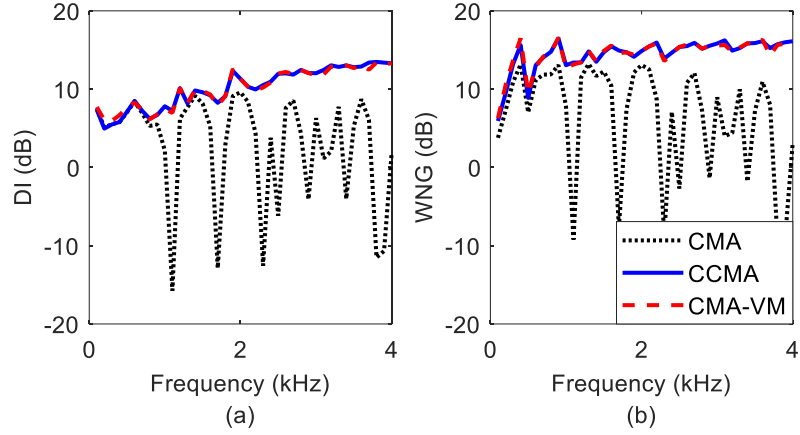

FIG. 6. (Color Online) Comparison of (a) the Directivity Index (DI) and (b) the White Noise Gain (WNG) for the CMA (black dotted) with $R = 0.12$ m and $M = 30$, the CCMA (blue solid) with $R_1 = 0.12$ m, $R_2 = 0.10$ m and $M_1 = M_2 = 30$, and the CMA-VM with $R_1 = 0.12$ m, $\tilde{R}_2 = 0.10$ m and $M_1 = \tilde{M}_2 = 30$.

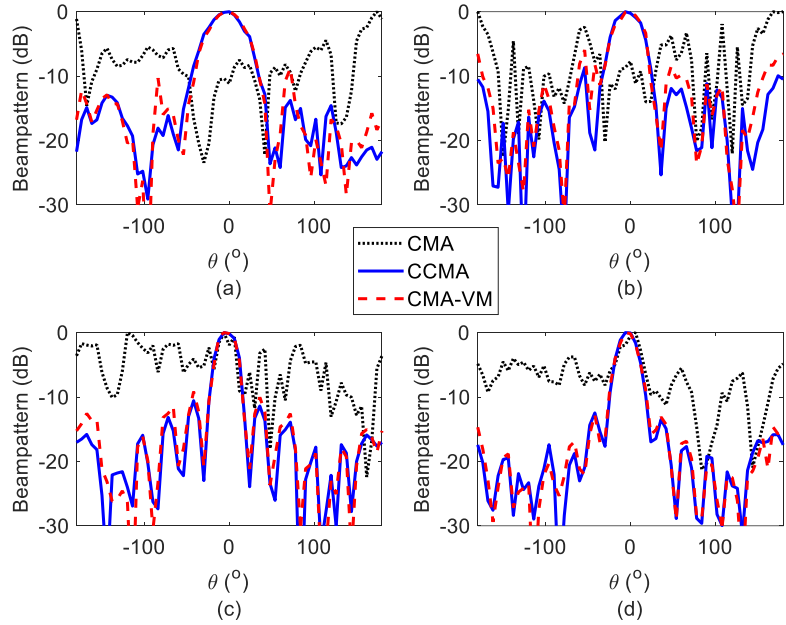

FIG. 7. (Color Online) Comparison of the beampattern at (a) $f = 1084$ Hz, (b) $f = 1728$ Hz, (c) $f = 2316$ Hz and (d) $f = 2490$ Hz for the CMA (black dotted) with $R = 0.12$ m and $M = 30$, the CCMA (blue solid) with $R_1 = 0.12$ m, $R_2 = 0.10$ m and $M_1 = M_2 = 30$, and the CMA-VM (red dashed) with $R_1 = 0.12$ m, $\tilde{R}_2 = 0.10$ m and $M_1 = \tilde{M}_2 = 30$.

## C. Results for 10 physical microphones

To further evaluate the performance of the proposed method, $M_1 = 10$ physical microphones are used in the CMA with a radius of $R_1 = 0.12$ m, whose beampattern is shown in FIG. 8(a). It is seen that, in addition to the performance degradation at the frequency corresponding to the zeros of the Bessel functions, spatial aliasing at higher frequencies is apparent. For a uniform CMA, to avoid spatial aliasing, the distance between microphones should be smaller than half the wavelength; therefore, the cutoff frequency is $f_{max} = c/(4R_1|\sin(\pi/M_1)|) = 2292$ Hz, in consistence with the beampattern in FIG. 8(a). Similarly, to remove the deep nulls in the DI and WNG of the CMA (shown by the black dotted curves in FIG. 9), an additional ring of $M_2 = 10$ physical microphones at a radius of $R_2 = 0.10$ m are added to form the CCMA. The beampattern of the CCMA is presented in FIG. 8(b), which shows that, although the spatial aliasing at the high frequency range still remains, the deep nulls due to the zeros of the Bessel functions are eliminated, as evidenced by the blue solid curves in FIG. 9.

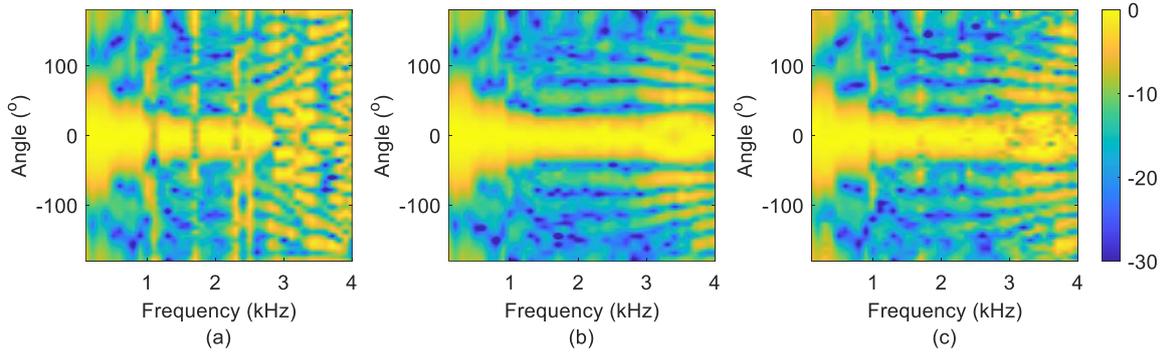

FIG. 8. (Color Online) Beampatterns for (a) the CMA with $R_1 = 0.12$ m and $M_1 = 10$, (b) the CCMA with $R_1 = 0.12$ m, $R_2 = 0.10$ m and $M_1 = M_2 = 10$, (c) the CMA-VM with $R_1 = 0.12$ m, $\tilde{R}_2 = 0.10$ m and $M_1 = \tilde{M}_2 = 10$.

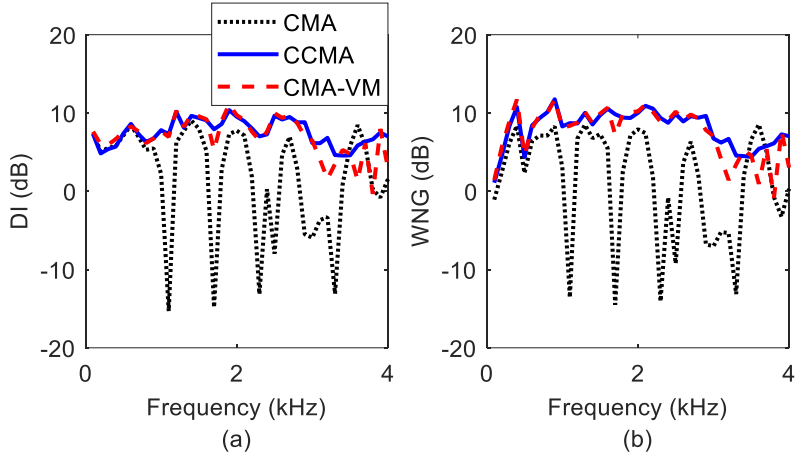

FIG. 9. (Color Online) Comparison of (a) the Directivity Index (DI) and (b) the White Noise Gain (WNG) for the CMA (black dotted) with $R = 0.12$ m and $M = 10$, the CCMA (blue solid) with $R_1 = 0.12$ m, $R_2 = 0.10$ m and $M_1 = M_2 = 10$, and the CMA-VM with $R_1 = 0.12$ m, $\tilde{R}_2 = 0.10$ m and $M_1 = \tilde{M}_2 = 10$.

By contrast, as shown by the beampattern in FIG. 8(c) and the DI and WNG in FIG. 9 (red dashed lines), the proposed CMA-VM, with $M_1 = 10$ physical microphones on the outer ring with a radius of $R_2 = 0.10$ m and $\tilde{M}_2 = 10$ virtual microphones on the inner ring with a radius of $\tilde{R}_2 = 0.10$ m, also mitigates the deep nulls problem in the CMA and restores the main lobes at these frequencies, although the side lobes are slightly higher than those of the CCMA. This can also be observed in FIG. 10 for the beampatterns at the first 4 null frequencies, which shows the side lobes of the CMA-VM (red dashed lines) are slightly higher than those of the CCMA (blue solid lines) but the main lobes are well reconstructed in comparison to the CMA (black dotted lines). It is noted that the result of the CMA-VM at the second null frequency ($f = 1728$ Hz) is much worse than that of the CCMA. As mentioned above, this may be attributed to some unknow measurement noise during the experiment and will be further investigated in the future.

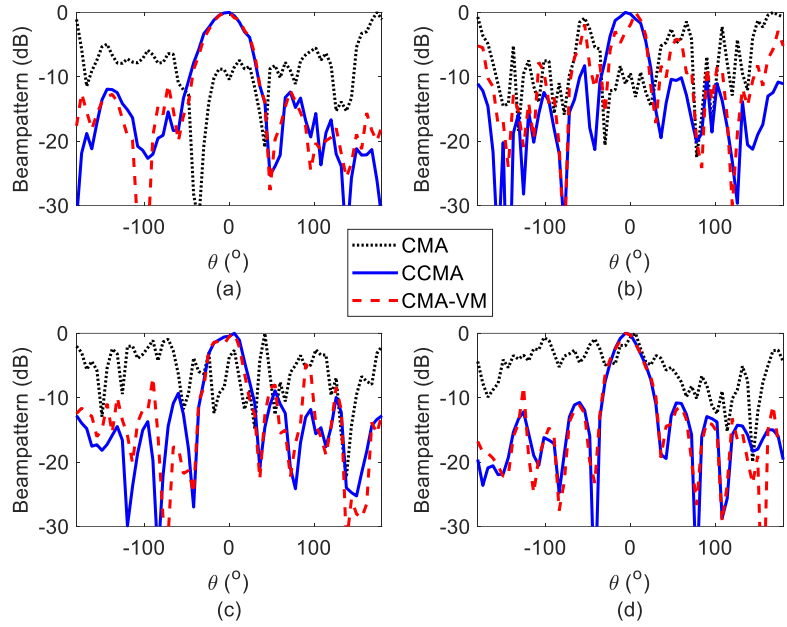

FIG. 10. (Color Online) Comparison of the beampatterns at (a) $f$ = 1084 Hz, (b) $f$ = 1728 Hz, (c) $f$ = 2316 Hz and (d) $f$ = 2490 Hz for the CMA (black dotted) with $R_1$ = 0.12 m and $M$ = 10, the CCMA (blue solid) with $R_1$ = 0.12 m, $R_2$ = 0.10 m and $M_1 = M_2$ = 10, and the CMA-VM (red dashed) with $R_1$ = 0.12 m, $\tilde{R}_2$ = 0.10 m and $M_1 = \tilde{M}_2$ = 30.

The above results demonstrated again that the proposed CMA-VM method can effectively mitigate the deep nulls in the DI and WNG in comparison to the CMA without the need for additional physical microphones, which is an advantage over the CCMA. However, it is noted that, both the CCMA and CMA-VM still suffer from significant spatial aliasing at higher frequencies above 3 kHz and the CMA-VM shows a performance drop in terms of DI and WNG in the high frequency range. To further investigate whether the proposed method can conquer the spatial aliasing effect by increasing the number of virtual microphones, experiments for three setups of the proposed CMA-VM method are carried out. For the first setup (CMA-VM-I) in FIG. 11(a), there are 10 physical microphones (black dots) on the outer ring with a radius of 0.12 m and 30 virtual microphones (green dots) on the inner ring with a radius of 0.10 m; for the second setup (CMA-VM-II) in FIG. 11(b), there are 10 physical and 20 virtual (30 in total) microphones on the outer ring and 10 virtual microphones on the inner ring; for the third setup (CMA-VM-III) in FIG. 11(c), there are 10 physical and 20 virtual (30 in total) microphones on the outer ring and 30 virtual microphones on the inner ring. It is noted that the sound pressures at all the virtual microphones in FIG. 11 are predicted from those measured by the physical microphones, no matter the virtual microphones are on the outer or inner rings. Therefore, for all the three setups in FIG. 11, only 10 physical microphones are used, as in the CMA whose beampattern is shown in FIG. 8(a).

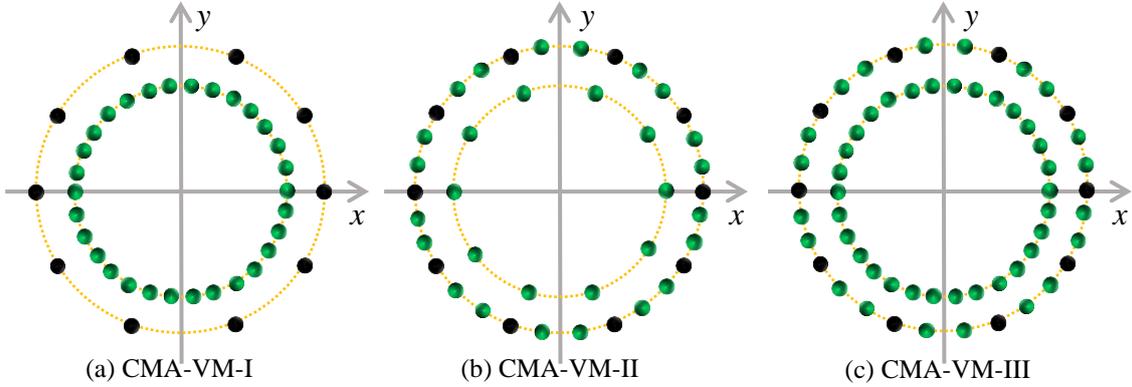

(a) CMA-VM-I      (b) CMA-VM-II      (c) CMA-VM-III

FIG. 11. (Color Online) Illustration of (a) CMA-VM-I with 10 physical microphones (black dots) on the outer ring with a radius of 0.12 m and 30 virtual microphones (green dots) on the inner ring with a radius of 0.10 m; (b) CMA-VM-II with 10 physical and 20 virtual (30 in total) microphones on the outer ring and 10 virtual microphones on the inner ring; and (c) CMA-VM-III with 10 physical and 20 virtual (30 in total) microphones on the outer ring and 30 virtual microphones on the inner ring.

The beampatterns for the three setups of the proposed CMA-VM approach are compared in FIG. 12, which shows the three setups to have similar beampatterns. In comparison to the beampatterns in FIG. 8, using more virtual microphones not only overcomes the deep-nulls problem in the CMA, but also eliminates the spatial aliasing at high frequencies. In addition, the main lobes are narrower because higher orders of circular modes can now be accommodated due to the increasing number of (virtual) microphones. On the other hand, by comparing the beampattern in FIG. 12(c), where 10 physical microphones are used, with the beampattern in FIG. 8(c), where 30 physical microphones are used, it can be seen that using more physical microphones can further improve the performance. This is reasonable because when more physical microphones are used, the sound pressures at the virtual microphones can be more accurately predicted by the AINN. Therefore, a trade-off between performance and the number of physical microphones must be balanced.

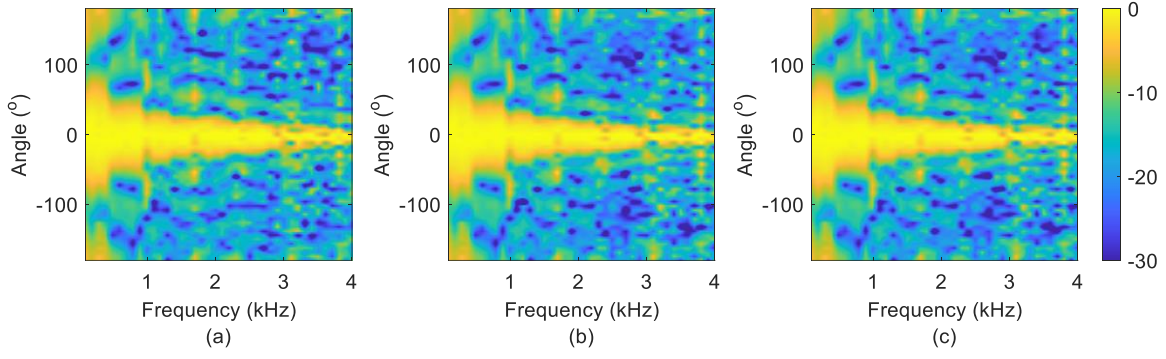

FIG. 12. (Color Online) Comparison of the beampatterns for (a) CMA-VM-I, (b) CMA-VM-II, and (c) CMA-VM-III.

The DI and WNG for the three setups of the proposed CMA-VM approach are compared in FIG. 13, which depicts that the DI is almost the same for the three setups, but the WNG increases with the number of (virtual) microphones. This is consistent with the previous finding that the WNG is proportional to the number of microphones used in the circular array (Huang et al., 2017a; Meyer, 2001). Both the DI and WNG in FIG. 13 are much higher than that in FIG. 9, showing a dramatic improvement in the beamformer performance by the proposed CMA-VM in comparison to the traditional CMA and even CCMA using more physical microphones. The above results fully prove that the proposed CMA-VM approach based on the AINN is superior to the CMA in terms of removing the deep nulls at frequencies corresponding to the zeros of Bessel functions and suppressing the spatial aliasing effect at high frequencies. It is worth emphasizing that these improvements are achieved without adding more physical microphones.

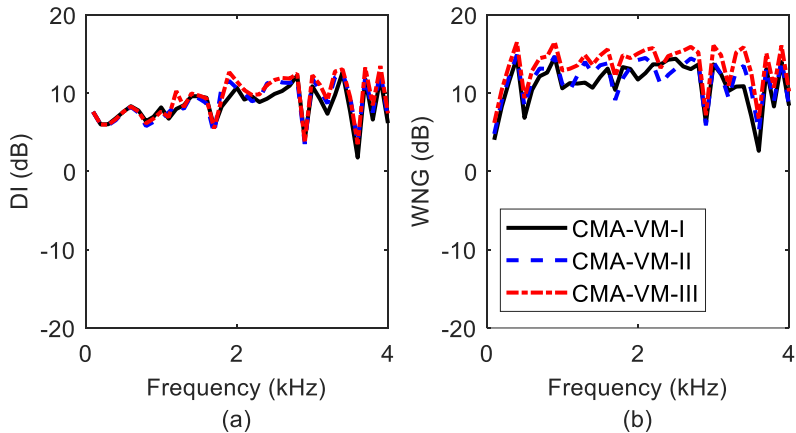

FIG. 13. (Color Online) Comparison of (a) DI and (b) WNG for CMA-VM-I, CMA-VM-II, and CMA-VM-III.

## V. CONCLUSIONS

This paper proposed a circular microphone array with virtual microphones to improve the performance of traditional circular microphone arrays at certain frequencies corresponding to the zeros of the Bessel functions yet without the need for more microphones to form a concentric circular array. The sound pressures at the virtual microphones were predicted from those measured by the physical microphones based on the acoustics-informed neural network. Experiments were carried out in a Hemi-Anechoic chamber to verify the performance of the proposed approach. Experimental results demonstrated that the proposed CMA-VM approach not only eliminates the deep nulls in the directivity index and the white noise gain, but also suppresses the spatial aliasing effect, thereby significantly improving the CMA's performance without increasing the number of microphones. Future work will investigate the effects of wall reflections and room reverberations on the performance of circular microphone arrays and the proposed approach.

## AUTHOR DECLARATIONS

The authors declare that there are no conflicts of interest to disclose.

## DATA AVAILABILITY

The data used in this paper is freely available online.

## REFERENCES

Benesty, J., Chen, J., and Cohen, I. (**2015**). *Design of Circular Differential Microphone Arrays*, Springer, Berlin Heidelberg, 172 pages.

Cai, S., Mao, Z., Wang, Z., Yin, M., and Karniadakis, G. E. (**2021a**). "Physics-informed neural networks (PINNs) for fluid mechanics: a review," Acta Mech. Sin., **37**, 1727–1738. doi:10.1007/s10409-021-01148-1

Cai, S., Wang, Z., Wang, S., Perdikaris, P., and Karniadakis, G. E. (**2021b**). "Physics-Informed Neural Networks for Heat Transfer Problems," J. Heat Transf., **143**, 060801. doi:10.1115/1.4050542

Huang, G., Benesty, J., and Chen, J. (**2017a**). "On the Design of Frequency-Invariant Beampatterns With Uniform Circular Microphone Arrays," IEEEACM Trans. Audio Speech Lang. Process., **25**, 1140–1153. doi:10.1109/TASLP.2017.2689681

Huang, G., Benesty, J., and Chen, J. (**2017b**). "Design of robust concentric circular differential microphone arrays," J. Acoust. Soc. Am., **141**, 3236–3249. doi:10.1121/1.4983122

Huang, G., Chen, J., and Benesty, J. (**2018**). "Insights into frequency-invariant beamforming with concentric circular microphone arrays," IEEEACM Trans. Audio Speech Lang. Process., **26**, 2305–2318. doi:10.1109/TASLP.2018.2862826

Huang, G., Cohen, I., Chen, J., and Benesty, J. (**2020**). "Continuously steerable differential beamformers with null constraints for circular microphone arrays," J. Acoust. Soc. Am., **148**, 1248–1258. doi:10.1121/10.0001770

Kahana, A., Papadimitropoulos, S., Turkel, E., and Batenkov, D. (**2023**). "A physically informed deep-learning approach for locating sources in a waveguide," J. Acoust. Soc. Am., **154**, 2553–2563. doi:10.1121/10.0021889

Karakonstantis, X., Caviedes-Nozal, D., Richard, A., and Fernandez-Grande, E. (**2024**). "Room impulse response reconstruction with physics-informed deep learning.," Retrieved from http://arxiv.org/abs/2401.01206

Karniadakis, G. E., Kevrekidis, I. G., Lu, L., Perdikaris, P., Wang, S., and Yang, L. (**2021**). "Physics-informed machine learning," Nat. Rev. Phys., **3**, 422–440. doi:10.1038/s42254-021-00314-5

Ma, F., Abhayapala, T. D., Samarasinghe, P. N., and Chen, X. (**2023**). "Spatial Upsampling of Head-Related Transfer Functions Using a Physics-Informed Neural Network.," Retrieved from http://arxiv.org/abs/2307.14650

Ma, F., Zhao, S., and Burnett, I. S. (**2024**). "Sound Field Reconstruction Using a Compact Acoustics-informed Neural Network.," doi:10.48550/arXiv.2402.08904

Meyer, J. (**2001**). "Beamforming for a circular microphone array mounted on spherically shaped objects," J. Acoust. Soc. Am., **109**, 185–193. doi:10.1121/1.1329616

Misyris, G. S., Venzke, A., and Chatzivasileiadis, S. (**2020**). "Physics-Informed Neural Networks for Power Systems," 2020 IEEE Power Energy Soc. Gen. Meet. PESGM, IEEE, Montreal, QC, Canada, 1–5. Presented at the 2020 IEEE Power & Energy Society General Meeting (PESGM). doi:10.1109/PESGM41954.2020.9282004

Nelson, P. A., and Elliott, S. J. (**1992**). *Active control of sound*, Academic Press, London, 436 pages.

Olivieri, M., Pezzoli, M., Antonacci, F., and Sarti, A. (**2021**). "A physics-informed neural network approach for nearfield acoustic holography," Sensors, , doi: 10.3390/s21237834. doi:10.3390/s21237834

Pavlidi, D., Griffin, A., Puigt, M., and Mouchtaris, A. (**2013**). "Real-Time Multiple Sound Source Localization and Counting Using a Circular Microphone Array," IEEE Trans. Audio Speech Lang. Process., **21**, 2193–2206. doi:10.1109/TASL.2013.2272524

Pezzoli, M., Antonacci, F., and Sarti, A. (**2023**). "Implicit neural representation with physics-informed neural networks for the reconstruction of the early part of room impulse responses.," doi:10.48550/arXiv.2306.11509

Teutsch, H., and Kellermann, W. (**2006**). "Acoustic source detection and localization based on wavefield decomposition using circular microphone arrays," J. Acoust. Soc. Am., **120**, 2724–2736. doi:10.1121/1.2346089

Tiana-Roig, E., Jacobsen, F., and Fernandez-Grande, E. (**2011**). "Beamforming with a circular array of microphones mounted on a rigid sphere (L)," J. Acoust. Soc. Am., **130**, 1095–1098. doi:10.1121/1.3621294

Tiana-Roig, E., Jacobsen, F., and Grande, E. F. (**2010**). "Beamforming with a circular microphone array for localization of environmental noise sources," J. Acoust. Soc. Am., **128**, 3535–3542. doi:10.1121/1.3500669

Torres, A. M., Cobos, M., Pueo, B., and Lopez, J. J. (**2012**). "Robust acoustic source localization based on modal beamforming and time–frequency processing using circular microphone arrays," J. Acoust. Soc. Am., **132**, 1511–1520. doi:10.1121/1.4740503

Wang, J., Yang, F., Li, J., Sun, H., and Yang, J. (**2023**). "Mode matching-based beamforming with frequency-wise truncation order for concentric circular differential microphone arrays," J. Acoust. Soc. Am., **154**, 3931–3940. doi:10.1121/10.0023964

Wang, J., Yang, F., Yan, Z., and Yang, J. (**2024**). "Design of frequency-invariant uniform concentric circular arrays with first-order directional microphones," Signal Process., **217**, 109330. doi:10.1016/j.sigpro.2023.109330

Wang, J., Yang, F., and Yang, J. (**2022**). "Insights Into the MMSE-Based Frequency-Invariant Beamformers for Uniform Circular Arrays," IEEE Signal Process. Lett., **29**, 2432–2436. doi:10.1109/LSP.2022.3224687

Zhao, S., Zhu, Q., Cheng, E., and Burnett, I. S. (**2022**). "A room impulse response database for multizone sound field reproduction (L)," J. Acoust. Soc. Am., **152**, 2505–2512. doi:10.1121/10.0014958